\documentclass[%
 reprint,
nofootinbib,
 amsmath,amssymb,
 aps,
]{revtex4-2}
\usepackage{braket}
\usepackage[export]{adjustbox}
\usepackage{lipsum}
\usepackage{soul}
\usepackage{subfigure}
\usepackage{lipsum}
\usepackage[colorlinks=true,linkcolor=blue]{hyperref}
\usepackage{amsmath}
\usepackage{mathrsfs}
\usepackage{mathtools, nccmath}
\usepackage{relsize}
\usepackage{tabularx}

\newcommand{\rrangle}{\rangle\! \rangle}
\newcommand{\llangle}{\langle\! \langle}

\makeatletter
\renewcommand*\env@matrix[1][\arraystretch]{%
  \edef\arraystretch{#1}%
  \hskip -\arraycolsep
  \let\@ifnextchar\new@ifnextchar
  \array{*\c@MaxMatrixCols c}}
\makeatother

\usepackage{cancel}
\usepackage{nicefrac}
\usepackage{graphicx}% Include figure files
\usepackage{dcolumn}% Align table columns on decimal point
\usepackage{bm}% bold math

\let\oldsqrt\sqrt
\def\sqrt{\mathpalette\DHLhksqrt}
\def\DHLhksqrt#1#2{%
\setbox0=\hbox{$#1\oldsqrt{#2\,}$}\dimen0=\ht0
\advance\dimen0-0.2\ht0
\setbox2=\hbox{\vrule height\ht0 depth -\dimen0}%
{\box0\lower0.4pt\box2}}

\begin{document}

\preprint{APS/123-QED}

\title{Second quantization of open quantum systems in Liouville space}% Force line breaks with \\

\author{Vladislav Sukharnikov}
\email{vladislav.sukharnikov@cfel.de}
\author{Stasis Chuchurka}%
\affiliation{
Deutsches Elektronen-Synchrotron DESY, Hamburg 22607, Germany\;
\\
Department of Physics, Universität Hamburg, Hamburg 22761, Germany
}%

\author{Andrei Benediktovitch}
\affiliation{
Deutsches Elektronen-Synchrotron DESY, Hamburg 22607, Germany
}%

\author{Nina Rohringer}
\email{nina.rohringer@desy.de}
\affiliation{
Deutsches Elektronen-Synchrotron DESY, Hamburg 22607, Germany\,
\\
Department of Physics, Universität Hamburg, Hamburg 22761, Germany
}%

\date{\today}

\begin{abstract}

We present a theoretical framework based on second quantization in Liouville space to treat open quantum systems. We consider an ensemble of identical quantum emitters characterized by a discrete set of quantum states. The second quantization is performed directly at the level of density matrices, thereby significantly reducing the size of the Liouville space. In contrast to conventional Hilbert space techniques, statistically mixed states and dissipation are naturally incorporated. As a particular example of application, we study the effect of incoherent processes and statistical mixing of emitters’ initial states in the interaction with quantum light. Moreover, we link our framework to a phase-space description of the dynamics, which can overcome the computational limitations of our method with the increasing number of particles.

\end{abstract}

\maketitle

\newpage

\section{\label{sec: introduction} Introduction}

The theoretical description of the static and dynamic properties of interacting macroscopic quantum systems with many degrees of freedom counts toward the biggest challenges in theoretical physics. A typically encountered problem is the interaction of an ensemble of quantum emitters (atoms, molecules, ions, etc.) with a quantized electromagnetic field. A realistic description of experimentally encountered systems further requires coupling to the environment, often treated as a reservoir with infinitely many degrees of freedom \cite{Breuer2007}. This coupling results in the decoherence of the density matrix associated with the system. For a macroscopic ensemble of quantum emitters, the direct solution of the master equation for the density matrix is a computationally intractable problem. Without introducing approximations or restrictions related to the symmetries of the system, the system's size scales exponentially with the number of involved emitters.

Imposing permutation invariance of the system, namely supposing that quantum emitters are indistinguishable, significantly reduces the number of involved degrees of freedom \cite{gegg2016efficient, gegg2017identical, gegg2017psiquasp, shammah2018open, bolanos2015algebraic, chase2008collective, xu2013simulating, hartmann2016generalized}. For such systems, the formalism of second quantization is a powerful theoretical framework \cite{landau1977} and is commonly applied to closed quantum systems. In open systems, the dissipation generally renders the Heisenberg picture inapplicable \cite{lemos1981heisenberg}. One needs to adopt the formalism of density matrices, with the dynamics taking place in Liouville space. In the limit of weak coupling to the environment, the density matrix satisfies the Lindblad-type master equation \cite{Breuer2007, lindblad1976generators, gorini1976completely, agarwal1974quantum, Manzano2020}. Assuming that all involved processes maintain permutation symmetry, the density matrix likewise preserves this symmetry in its temporal evolution \cite{shammah2017superradiance, shammah2018open, hartmann2016generalized, chase2008collective}. Under this assumption, the concept of second quantization can be extended to density matrices.

Only a relatively small body of literature targets the second quantization formalism in Liouville space, despite this method's potential impact for modeling dissipative systems. Several other approaches take advantage of permutation symmetry. The general Dicke basis \cite{chase2008collective, hartmann2016generalized, shammah2017superradiance, shammah2018open} was used for the efficient treatment of dissipative systems of identical two-level particles. Its extension to many levels, however, is nontrivial. Recently, the evolution of identical multi-level particles was treated in a second quantization picture \cite{Silva2022}, assuming a collective interaction with the environment, which restricts the Hilbert space to fully symmetric states. When particles interact with the environment independently, the system's quantum state does not necessarily evolve in the subspace of fully symmetric states, therefore the method's applicability is limited. In Refs. \cite{gegg2016efficient, gegg2017identical, gegg2017psiquasp, bolanos2015algebraic}, master equations with non-collective dissipation are solved by introducing an occupation-number representation for density matrices. This representation is closely related to the Jordan-Schwinger bosonization in the Liouville space \cite{bolanos2015algebraic, AlvarezGiron2020}.

For a quantum system described by second quantization in Hilbert space, a direct extension of bosonic and fermionic operators to the Liouville space is possible by doubling the number of degrees of freedom \cite{schmutz1978real, prosen2008third, harbola2008superoperator, dzhioev2012nonequilibrium}. This procedure implies certain symmetry restrictions and accounts for incoherent processes only in the form of particle loss or collective dissipation. Second quantization formalism for treating incoherent processes in a system with a fixed number of particles is still lacking.

In this work, we formulate the second quantization in the Liouville space for open quantum systems consisting of a fixed number of identical particles. In contrast to Ref. \cite{Silva2022}, we are not restricted to the collective interaction with the environment and can also treat a wider class of statistically mixed states. We show that the dynamics of $N$ identical $M$-level particles is formally equivalent to $M^2$ harmonic oscillators that share $N$ excitations. To illustrate the formalism's applicability, we focus our discussion on light-matter interaction. In particular, we revisit the phenomenon of collective spontaneous emission from a macroscopic ensemble of atoms --- superradiance and superfluorescence \cite{Benedict2018, dicke1954coherence, gross1982superradiance, agarwal1974quantum, carmichael1999statistical}. We show conservation laws at the level of occupation numbers for density matrices, further reducing the number of degrees of freedom. The density matrix evolves as a sum of independent contributions with specific values of the integrals of motion, making the numerical simulation parallelizable.

The article is organized as follows. Section \ref{sec:II} introduces the framework of second quantization for density matrices by defining a properly extended Hilbert space isomorphic to the Liouville space. In Sec. \ref{sec:III}, we provide a specific discussion on the second quantization representation of density matrices and observables. In Sec. \ref{sec:IV.A}, we consider a particular system of $N$ identical two-level emitters coupled to a common environment and investigate the effect of statistical mixing of the initial state on collective spontaneous emission. Section \ref{sec:IV.B} describes the dynamics of an incoherently pumped $\Lambda$ system with dissipation. In Sec. \ref{sec:IV.C}, the effect of statistical mixing is studied in the Tavis-Cummings model. A connection of our formalism to phase-space techniques is discussed in Sec. \ref{sec:V}.

%%%%%%%%%%%%%%%%%%%%%%%%%%%%%%%%%%%%%%%%%%%%%%%%%
\section{\label{sec:II} General Formalism of second quantization}
%%%%%%%%%%%%%%%%%%%%%%%%%%%%%%%%%%%%%%%%%%%%%%%%%

In this section, we extend the traditional second quantization technique, formulated for state vectors, to density matrices that belong to the Liouville space. We impose only two conditions on the system: permutation symmetry and conservation of the number of particles.

%%%%%%%%%%%%%%%%%%%%%%%%%%%%%%%%%%%%%%%%%%%%%%%%%
\subsection{Hilbert space of identical particles}\label{sec:II.A}
%%%%%%%%%%%%%%%%%%%%%%%%%%%%%%%%%%%%%%%%%%%%%%%%%

Without loss of generality, we consider an ensemble of $N$ atoms, each characterized by the same set of $M$ electronic states, interacting with a quantized radiation field. The spatial size of the sample is much smaller than any relevant wavelength of radiation, i.e. of either external field or related to radiative electronic transitions. This translates into the invariance of the Hamiltonian under rearrangements of atoms.

The field-free eigenstates of the Hamiltonian of atom $\mu$ fulfill
\begin{equation*}
     \widehat{H}_{\mu} \, |i\rangle_\mu = \hbar \omega_i \, |i\rangle_\mu \quad (i = 1, \ldots, M),
\end{equation*}
with eigenvalue $\hbar \omega_i$ pertaining to the eigenstate $|i\rangle$. A state of the whole ensemble can be expanded in terms of these eigenstates:
\begin{equation}\label{eq:psi}
    |\psi \rangle = \sum_{\pmb{i} } C_{\pmb{i}} \; |i_1\rangle_{1} \; |i_2\rangle_{2} \ldots |i_N\rangle_{N},
\end{equation}
where the summation runs over the multi-index $\pmb{i} = (i_1, \ldots, i_N)$, $i_\mu = 1, \ldots, M$. The coefficients $C_{\pmb{i}}$ ensure that this state is permutation invariant. The following projection operators are crucial for our analysis:
\begin{equation}\label{eq:sigma_operators}
    \widehat{\sigma}_{\mu,ij}= |i\rangle_\mu \, \langle j|_\mu.
\end{equation}

Permutation-invariant states of $N$ particles can be constructed by appropriately defined creation operators, a prominent example being the Schwinger representation \cite{schwinger1965quantum, agarwal1970master, klein1991boson, kumar1980theory, biedenharn1984angular, carusotto1989dynamics, drobny1992quantum, wu1999schwinger, walls1970quantum}. We provide a brief overview of the Schwinger representation for $M$-level atoms in Appendix \ref{Appendix:A} and set the goal to extend this concept to density matrices. 

%%%%%%%%%%%%%%%%%%%%%%%%%%%%%%%%%%%%%%%%%%%%%%%%%%%%%%%%%%%%%%%%%%%%%%%%
\subsection{Liouville space of identical particles}\label{sec:II.B}
%%%%%%%%%%%%%%%%%%%%%%%%%%%%%%%%%%%%%%%%%%%%%%%%%%%%%%%%%%%%%%%%%%%%%%%%

The state-vector representation is inappropriate when describing open quantum systems. Such situations require a more general framework of density matrices, which are elements of the Liouville space. Under the condition that coupling to the environment by any incoherent processes preserves the permutation symmetry, it is possible to introduce the concept of second quantization in Liouville space.

The density matrix for a pure state \eqref{eq:psi} can be represented in terms of $\sigma$-operators \eqref{eq:sigma_operators}:
\begin{equation*}
    \widehat{\rho} = |\psi\rangle \langle \psi|= \sum_{\pmb{ij}} C_{\pmb{i}} \, C_{\pmb{j}}^* \;\, \widehat{\sigma}_{1, i_1 j_1} \ldots \widehat{\sigma}_{N, i_N j_N}.
\end{equation*}
When interaction with an environment is included, the composite density matrix has a different form. 
\begin{equation*}
    \widehat{\rho}_\text{\,full} = \sum_{\mathclap{ \pmb{ij}, \text{env}, \text{env}' } } \, C_{\pmb{i}, \text{env}} \, C_{\pmb{j},\text{env}'}^* \;\, \widehat{\sigma}_{1, i_1 j_1} \ldots \widehat{\sigma}_{N, i_N j_N} \; |\text{env} \rangle \langle \text{env}'|,
\end{equation*}
where $\text{env}, \text{env}'$ represent the environment's degrees of freedom.

In most applications, the evolution of the environment has no significance and is neglected by averaging over its degrees of freedom. This average changes the structure of the system's reduced density matrix:
\begin{equation}\label{eq:density_matrix_sigma}
    \widehat{\rho} = \mathrm{Tr}_{\,\text{env}}\, \widehat{\rho}_\text{\,full} = \sum_{\pmb{ij}} \overline{C
    _{\pmb{ij}}} \;\, \widehat{\sigma}_{1, i_1 j_1} \ldots \widehat{\sigma}_{N, i_N j_N},
\end{equation}
with $\overline{C
    _{\pmb{ij}}} = \sum_{\text{env}} C_{\pmb{i},\text{env}} \, C^*_{\pmb{j},\text{env}}$. In general, the resulting density matrix describes a statistical mixture, and the traditional second quantization is no longer applicable despite the preserved permutation symmetry.

The structure of \eqref{eq:density_matrix_sigma} resembles the state vector \eqref{eq:psi}. Following Refs. \cite{fano1964liouville, bolanos2015algebraic, gyamfi2020fundamentals, Manzano2020}, we treat the density matrix as a state vector in an extended Hilbert space $\mathscr{H}$ we call \textit{Liouville-Hilbert} space. The building blocks of this space,
\begin{equation}\label{eq:superket}
     \widehat\sigma_{\mu,ij} \equiv |ij\rrangle_\mu,
\end{equation}
are called \textit{superkets}. The density matrices are isomorphic to vectors in this extended Hilbert space:
\begin{equation}\label{eq:density_matrix}
    \widehat{\rho} = |\rho\rrangle = \sum_{\pmb{ij}} \overline{C
    _{\pmb{ij}}} \; |i_1 j_1\rrangle_1 \ldots |i_N j_N \rrangle_N.
\end{equation}
A supervector notation for the density matrix $|\rho\rrangle$ highlights its belonging to the Liouville-Hilbert space.

In general, there are $M^{2N}$ terms in Eq.\ \eqref{eq:density_matrix}, reflecting the exponential wall of the Liouville space with the increasing number of particles $N$. Recalling our postulate that particles are indistinguishable, we restrict our considerations solely to the subspace of \textit{symmetric} density matrices. The size of this subspace is polynomial in $N$ \cite{gegg2016efficient, gegg2017identical, gegg2017psiquasp, shammah2018open, bolanos2015algebraic, chase2008collective, xu2013simulating, hartmann2016generalized}, namely
\begin{equation}\label{eq:size}
 \begin{pmatrix}
        N + M^2 -1\\
        N
    \end{pmatrix} 
    = \dfrac{(N + M^2 - 1)!}{N! \, (M^2 - 1)!}.
\end{equation}

In contrast to the state vectors of eq. \eqref{eq:psi}, which are either symmetric or antisymmetric under the particle exchange, the corresponding density matrices appear to be only \textit{symmetric}. Thus, we do not specify the statistical properties of particles and keep the description as general as possible.

%%%%%%%%%%%%%%%%%%%%%%%%%%%%%%%%%%%%%%%%%%%%%%%%%
\subsection{Second quantization in Liouville-Hilbert space}\label{sec:II.C}
%%%%%%%%%%%%%%%%%%%%%%%%%%%%%%%%%%%%%%%%%%%%%%%%%

To adapt the Schwinger bosonization for density matrices, $\mathscr{H}$-space requires a properly defined inner product. For this purpose, a dual space $\mathscr{H}^*$ spanned by \textit{superbras} $\llangle ij|$ is introduced \cite{gyamfi2020fundamentals}, satisfying $\llangle ij | pq \rrangle = \delta_{ip} \, \delta_{jq}$. In general, the inner product is given by:
\begin{equation}\label{eq:scalar_prod}
    \llangle A | B \rrangle
    =
    \mathrm{Tr} \big\{ \widehat{A}^{\, \dagger} \widehat{B} \big\}.
\end{equation}

Appendix \ref{App:B.1} contains a brief algebraic derivation of the following bosonization rule:
\begin{equation}\label{eq:sigma_L_bosonization}
    \sum_{\mu = 1}^{N}\, \widehat{\sigma}_{\mu,pq} \, \widehat{\rho}
    = 
    \sum_{t = 1}^{M}\, \widehat{b}_{pt}^{\,\dagger} \, \widehat{b}_{qt} \, |\rho\rrangle.
\end{equation}
Here, we have introduced \textit{collective bosonic operators} with double subscripts that satisfy the commutation relations 
\begin{equation*}
    \big[\hspace{0.35mm} \widehat{b}_{ij}, \, \widehat{b}_{st}^{\,\dagger} \big] 
    =
    \delta_{is} \, \delta_{jt}.
\end{equation*}
These operators change the number of particles occupying generalized states \eqref{eq:superket}. Before defining the generalized occupation-number basis, we address a few other essential points related to the bosonization of operators.

%%%%%%%%%%%%%%%%%%%%%%%%%%%%%%%%%%%%%%%%%%%%%%%%%
\subsubsection*{Quantization of operators acting from the right}
%%%%%%%%%%%%%%%%%%%%%%%%%%%%%%%%%%%%%%%%%%%%%%%%%

When the density matrix is treated as an operator, the action upon it from the right is a well-defined operation. We denote the right action of any superoperator with the superscript $^\intercal$:
\begin{equation*}
    \widehat{\mathcal{O} }^{\,\intercal} \widehat{\rho} \equiv \widehat{\rho}\; \widehat{\mathcal{O}}.
\end{equation*}

The bosonization of the right-acting collective operator is derived in Appendix \ref{App:B.2}:
\begin{equation}\label{eq:sigma_R_bosonization}
  \sum_{\mu = 1}^{N} \widehat{\sigma}^{\,\intercal}_{\mu,k\ell} \, \widehat{\rho} 
  \equiv
   \sum_{\mu = 1}^{N} \widehat{\rho} \; \widehat{\sigma}_{\mu,k\ell} 
     = 
     \sum_{s = 1}^{M}\, \widehat{b}_{s\ell}^{\,\dagger} \, \widehat{b}_{sk} \, |\rho\rrangle,
\end{equation}
which resembles the expression \eqref{eq:sigma_L_bosonization}, but with a summation running over the first subscript $s$. As a result of the bosonization, the right action of the collective operator amounts to the \textit{left action} of bosonic superoperators.

In addition to Eqs.\ \eqref{eq:sigma_L_bosonization} and \eqref{eq:sigma_R_bosonization}, the following terms are encountered in master equations describing atoms independently coupled to a reservoir \cite{gegg2017identical}:
\begin{equation}\label{eq:lindlad_superoperators}
    \widehat{\Gamma}_{pq}^{\, k\ell}
    =
    \sum_{\mu = 1}^{N} \widehat{\sigma}_{\mu,pq} \, \widehat{\sigma}^{\,\intercal}_{\mu,k\ell}
    =
    \widehat{b}_{p\ell}^{\,\dagger} \, \widehat{b}_{qk}.
\end{equation}
Within the traditional framework of second quantization, such terms cannot be bosonized.

%%%%%%%%%%%%%%%%%%%%%%%%%%%%%%%%%%%%%%%%%%%%%%%%%
\subsubsection*{Many-particle permutation invariant operators}
%%%%%%%%%%%%%%%%%%%%%%%%%%%%%%%%%%%%%%%%%%%%%%%%%

Many-particle operators are quantized in the same fashion. For example, a two-particle permutationally invariant operator,
\begin{equation*}
    \widehat{V}^{(2)}
    =
    \sum_{ \mathclap{ \mu_1 \not = \mu_2 } } \, \widehat{V}_{\mu_1 \mu_2},
\end{equation*}
is quantized as follows:
\begin{multline}\label{eq:two_body_quantization}
        \widehat{V}^{(2)}
        =
        \sum_{ \mathclap{\bm{stij} } } \, \llangle s_1 t_1|_{1} \, \llangle s_2 t_2|_{2} \, \widehat{V}_{12} \, |i_2 j_2 \rrangle_{2}\, |i_1 j_1 \rrangle_{1} 
        \\
        \times \, \widehat{b}_{s_1 t_1}^{\, \dagger} \, \widehat{b}^{\, \dagger}_{s_2 t_2} \, \widehat{b}_{i_2 j_2} \, \widehat{b}_{i_1 j_1},
\end{multline}
where the matrix element is the same for each pair of particles.

%%%%%%%%%%%%%%%%%%%%%%%%%%%%%%%%%%%%%%%%%%%%%%%%%
\subsection{Occupation-number basis}\label{sec:II.D}
%%%%%%%%%%%%%%%%%%%%%%%%%%%%%%%%%%%%%%%%%%%%%%%%%

The remaining step is to introduce the occupation-number basis, i.e., the generalized Fock states:
\begin{equation*}
    |\{n_{ij}\} \rrangle
    =
    |\{n_{11}, n_{12}, \ldots, n_{1M}, \ldots, n_{MM}\}\rrangle,
\end{equation*}
where $\{n_{ij}\}$ denotes an ordered sequence of $M^2$ numbers. The elements of this basis are orthonormal with respect to the inner product \eqref{eq:scalar_prod}. However, this inner product has no direct physical meaning and was only required to expand the Liouville space to Liouville-Hilbert space$\,\,\mathscr{H}$. For calculating expectation values, one has to transfer trace operator$\,\,\mathrm{Tr}$ from the Liouville space, where its operation on matrices is evident, to the Liouville-Hilbert space of supervectors.

Vectors $|\{n_{ij}\} \rrangle$ are symmetrized combinations of $\sigma$ operators. Their trace is zero if states $|i\not = j\rrangle$ are occupied, i.e., there are nonzero numbers with different indices $n_{i\not = j}$. In the opposite case, when all $n_{i\not = j}=0$, the trace results in
\begin{equation}\label{eq:trace_of_a_basis_vector}
    \text{Tr} \, | \{ n_{ii} \} \rrangle
    =
    \sqrt{\frac{N!}{ \prod_{q} n_{qq}! } }.
\end{equation}
The radicand is the number of ways to distribute particles over the occupied states.

The bosonic operators $\widehat{b}_{ij}$ and $\widehat{b}^{\,\dagger}_{ij}$ annihilate and create particles in a state defined by a pair of subscripts:
\begin{align*}
    & \widehat{b}_{ij} \; |\{\ldots, n_{ij}, \ldots\} \rrangle 
    =
    \sqrt{n_{ij}} \; |\{\ldots, n_{ij} - 1, \ldots\} \rrangle,
    \\
    & \widehat{b}^{\,\dagger}_{ij} \; |\{\ldots, n_{ij}, \ldots\} \rrangle
    =
    \sqrt{n_{ij} + 1} \; |\{\ldots, n_{ij} + 1, \ldots\} \rrangle.
\end{align*}
Since the number of particles $N$ is conserved, the bosonic operators appear in product pairs containing the same number of creation and annihilation operators, hence the operator identity:
\begin{equation}\label{eq:fixed_number}
    \sum_{ \mathclap{ p, q = 1 } }^{M} \, \widehat{b}^{\, \dagger}_{pq} \, \widehat{b}_{pq}
    =
    \sum_{ \mathclap{ p, q = 1 } }^{M} \, \widehat{n}_{pq}
    \equiv
    N.
\end{equation}
In that sense, the ensemble of $N$ identical $M$-level particles is equivalent to $M^2$ harmonic oscillators that share $N$ excitations. 

Formally, the generalized Fock states are built from a vacuum superstate $|\mathrm{vac}\rrangle$, in complete analogy to Eq.\,\,\eqref{eq:HS_occupation_numbers} of Appendix \ref{Appendix:A}:
\begin{equation*}
    |\{n_{ij}\} \rrangle
    =
    \mathlarger{\prod}_{ \mathclap{ p, q = 1 } }^M \, \dfrac{ \big( \, \widehat{b}_{pq}^{\,\dagger} \big)^{n_{pq}} }{\sqrt{n_{pq}!}} \, |\mathrm{vac} \rrangle.
\end{equation*}
The vacuum superstate does not describe any physical system and serves only as a foundation for constructing physical states. 

A similar but conceptually different Liouville-Fock space appears in Refs. \cite{schmutz1978real, prosen2008third, harbola2008superoperator, dzhioev2012nonequilibrium} for already second-quantized bosonic and fermionic systems. In contrast, we do not rely on the preceding second quantization in Hilbert space since we perform it directly in the Liouville-Hilbert space.

\begin{table*}[hbt!]
    \begin{tabular}{ c||c||c||c }
        Form of & Pure uncorrelated & Mixed uncorrelated & Symmetric Dicke states \\
        representation & quantum state & quantum state & (two-level atoms)
        \\ [0.2em] \hline
        \rule{0pt}{20pt}
        State vector
        &
        $\displaystyle |\psi\rangle = \bigotimes\limits_{\mu = 1}^{N} \left\{\sum_{i = 1}^{M} c_i\, |i\rangle_\mu \right\}$
        &
        ---
        &
        $|L_N\rangle$, $L$ atoms out of $N$ are excited
        \\[1.5em]
        \hline
        \rule{0pt}{20pt}
        \; Density matrix \;
        &
        $\displaystyle \widehat{\rho} = \prod\limits_{\mu = 1}^{N} \left\{\sum_{i, j = 1}^{M} c_i \, c_j^* \; \widehat{\sigma}_{\mu, ij} \right\}$ 
        &
        $\displaystyle \widehat{\rho} = \prod\limits_{\mu = 1}^{N} \left\{\sum_{i = 1}^{M} p_i\; \widehat{\sigma}_{\mu, ii} \right\}$ 
        &
        $\displaystyle \widehat{\rho}_{L} = |L_N\rangle \langle L_N|$
        \\[1.5em]
        &
        $\displaystyle \sum\limits_{i=1}^{M} |c_i|^2 = 1$
        &
        $\displaystyle \sum\limits_{i=1}^{M} p_i = 1, \; p_i > 0$
        &
        $L \in [0,N]$
        \\[1.5em]\hline
        \rule{0pt}{20pt}
        Occupation number \;
        & 
        \; $\displaystyle \rho(\{n_{ij}\}) = N! \, \prod_{\mathclap{p, q = 1}}^{M} \dfrac{\left( c_p \, c_q^* \right)^{n_{pq}}}{n_{pq}!}$\;
        &
        \;$\displaystyle \rho(\{n_{ii}\}) = N!\, \prod_{\mathclap{q = 1}}^{M} \dfrac{(p_q)^{n_{qq}}}{n_{qq}!} $\;
        &
        \; $\displaystyle \rho_{L}(\{n_{ij}\}) = \dfrac{(N - L)! \, L! }{n_{11}!\, n_{12}!\, n_{21}!\, n_{22}!}$ \;
        \\[1.5em]
        representation \eqref{eq:density_matrix_n}
        &

        &
        all $\displaystyle n_{i \not = j} = 0$
        &
        \; $n_{12} = n_{21}$, $n_{22} - n_{11} = 2L - N$ \;
    \end{tabular}
\caption{Different quantum states, rewritten in the occupation-number representation \eqref{eq:density_matrix_n}. Columns represent various quantum states: uncorrelated pure, uncorrelated mixed states, and Dicke states. In the first row, we provide their state representation in the Hilbert space (where applicable), followed by their density matrix in the second row. The last row contains the occupation-number representation of the given states.}
\label{table:1}
\end{table*}

The possibility of introducing occupation numbers has been noted in previous approaches. In Ref. \cite{gegg2017identical}, the change of occupation numbers was implemented with the help of superoperators \eqref{eq:lindlad_superoperators} with the algebra:
\begin{equation*}
    \left[\widehat{\Gamma}_{mn}^{\, op},\, \widehat{\Gamma}_{qr}^{\, st} \right] 
    =
    \widehat{\Gamma}_{mr}^{\, sp} \, \delta_{qn} \, \delta_{ot} - \widehat{\Gamma}_{qn}^{\, ot} \, \delta_{mr} \, \delta_{sp}.
\end{equation*}
Bosonized representation of these operators \eqref{eq:lindlad_superoperators} automatically preserves their algebra. Therefore, bosonic operators with double subscripts are Jordan-Schwinger operators, as also pointed out in Ref. \cite{bolanos2015algebraic}.

%%%%%%%%%%%%%%%%%%%%%%%%%%%%%%%%%%%%%%%%%%%%%%%%%
\section{Second quantization representation}\label{sec:III}
%%%%%%%%%%%%%%%%%%%%%%%%%%%%%%%%%%%%%%%%%%%%%%%%%

In this chapter, we draw the general picture of applying our formalism by discussing crucial steps: representation of density matrices in the occupation-number basis, time evolution, and calculation of expectation values of observables.

%%%%%%%%%%%%%%%%%%%%%%%%%%%%%%%%%%%%%%%%%%%%%%%%%
\subsection{Density matrices}\label{sec:III.A}
%%%%%%%%%%%%%%%%%%%%%%%%%%%%%%%%%%%%%%%%%%%%%%%%%

The first step is expanding the density matrix in the occupation-number basis as:
\begin{equation}\label{eq:density_matrix_n}
    |\rho\rrangle
    =
    \mathlarger{\sum}_{\mathclap{\{n_{ij}\}}} \; \sqrt{\dfrac{\prod_{pq} n_{pq}!}{N!}} \, \rho\left(\{n_{ij}\}\right) \, |\{n_{ij}\} \rrangle.
\end{equation}
Since the number of atoms $N$ is fixed, the expansion coefficients $\rho(\{n_{ij}\})$ are nonzero only if the sum of occupation numbers is $N$, leaving at most \eqref{eq:size} non-trivial terms in the summation. The hermiticity of the density matrix leads to condition $\rho(\{n_{ij}\}) = \rho^*(\{n_{ji}\})$, further reducing the number of nontrivial terms.

The multinomial normalization in \eqref{eq:density_matrix_n} simplifies the expressions for the expectation values by compensating the trace \eqref{eq:trace_of_a_basis_vector}. The trace of the density matrix simply becomes:
\begin{equation*}
    \mathrm{Tr} \, |\rho \rrangle
    =
    \mathlarger{\sum}_{\mathclap{ \substack{ \{n_{ij}\} \\ n_{i\not = j} = 0}}} \, \rho(\{n_{ij}\}) = 1,
\end{equation*}
where the sum is constrained to $n_{i\neq j} = 0$. 

Table \ref{table:1} gives a few examples of quantum states rewritten in the form \eqref{eq:density_matrix_n}, among them are uncorrelated symmetric pure and mixed states with a tensor product structure. The expression for Dicke states of two-level atoms is also given. As already mentioned, density matrices for antisymmetric states are \textit{symmetric} themselves, meaning that fermionic states can be built using bosonic superoperators. The fermionic density matrix has a structure of a determinant built from the creation superoperators acting on the vacuum superstate:
 \begin{equation*}
    |\rho \rrangle = \dfrac{1}{\sqrt{N!}}\, 
    \begin{vmatrix}[2]
        \widehat{b}_{i_1 i_1}^{\,\dagger} & \widehat{b}_{i_1 i_2}^{\,\dagger} & \ldots & \widehat{b}_{i_1 i_N}^{\,\dagger}
        \\
        \widehat{b}_{i_2 i_1}^{\,\dagger} & \widehat{b}_{i_2 i_2}^{\,\dagger} & \ldots & \widehat{b}_{i_2 i_N}^{\,\dagger}
        \\
        \vdots & \vdots & \ddots & \vdots \\
        \, \widehat{b}_{i_N i_1}^{\,\dagger} & \widehat{b}_{i_N i_2}^{\,\dagger} & \ldots & \widehat{b}_{i_N i_N}^{\,\dagger}
    \end{vmatrix} 
    |\mathrm{vac} \rrangle.
\end{equation*}
The corresponding state is antisymmetrized $N$-partite vector $|i_1\rangle \, |i_2\rangle \ldots |i_N\rangle$.

%%%%%%%%%%%%%%%%%%%%%%%%%%%%%%%%%%%%%%%%%%%%%%%%%
\subsection{Observables and expectation values}\label{sec:III.B}
%%%%%%%%%%%%%%%%%%%%%%%%%%%%%%%%%%%%%%%%%%%%%%%%%

The expectation value $\langle \mathcal{O} \rangle$ of the permutationally invariant operator $\widehat{\mathcal{O}}$ is the following trace:
\begin{equation*}
    \langle \mathcal{O} \rangle
    =
    \mathrm{Tr} \big\{ \widehat{\mathcal{O}} \, \widehat{\rho} \big\}.
\end{equation*}
A straightforward way to find this expectation value is bosonizing the operator following Eqs.\ \eqref{eq:sigma_L_bosonization}--\eqref{eq:two_body_quantization}, acting on the density matrix expanded as \eqref{eq:density_matrix_n}, and finally making use of \eqref{eq:trace_of_a_basis_vector}.

Alternatively, one may interpret the expectation value as the projection of the system's density matrix onto some supervector $|\mathcal{O}\rrangle$ representing the operator $\widehat{\mathcal{O}}$ \cite{gyamfi2020fundamentals}:
\begin{equation*}
    \langle \mathcal{O} \rangle
    =
    \mathrm{Tr} \big\{ \widehat{\rho}^{\,\dagger} \, \widehat{\mathcal{O}} \big\}
    =
    \llangle \rho | \mathcal{O} \rrangle.
\end{equation*}
Being the combination of $\sigma$ operators, permutationally invariant operators are isomorphic to symmetric supervectors belonging to the Liouville-Hilbert space and thus can be expressed in the occupation-number representation:
\begin{equation*}
    |\mathcal{O}\rrangle
    =
    \mathlarger{\sum}_{\{n_{ij}\}} \, \sqrt{ \dfrac{N!}{ \prod_{pq} n_{pq}! } } \, \mathcal{O}( \{n_{ij} \} ) \; |\{n_{ij}\} \rrangle.
\end{equation*}
Here, similarly to $\rho(\{n_{ij}\})$, $\mathcal{O}(\{n_{ij} \} )\,$ is nonzero only if the sum of occupation numbers is $N$. Note that the normalizing coefficient for operators is inverse to that of \eqref{eq:density_matrix_n}. The expectation value in the state \eqref{eq:density_matrix_n} is then easily calculated by
\begin{equation}\label{eq:average}
    \langle \mathcal{O} \rangle
    =
    \sum_{ \mathclap{ \{n_{ij}\} } } \; \rho(\{n_{ji}\}) \; \mathcal{O}(\{n_{ij}\}).
\end{equation}
In practical application, the major difficulty is finding these coefficients. The derivation of coefficients $\mathcal{O}(\{n_{ij} \} )\,$ for a given operator $\widehat{\mathcal{O}}$ can be found in Appendix \ref{appendix:generating_functional}. Here, we demonstrate how to apply the derived expressions.

A permutationally invariant $K$-particle operator can be decomposed in terms of $\sigma$-operators:
\begin{equation}
\label{eq:K_body_operator}
    \widehat{\mathcal{O}}^{(K)}
    =
    \sum\limits_{\pmb{ij}} \mathcal{O}^{(K)}_{\pmb{ij}} \; \mathlarger{\sum}\limits_{\mathclap{\mu_1 \not = \ldots \not = \mu_K}}\; \widehat{\sigma}_{\mu_1, i_1 j_1} \ldots \widehat{\sigma}_{\mu_K, i_K j_K}.
\end{equation}
The matrix elements $\mathcal{O}^{(K)}_{\pmb{ij}}$ are directly used for calculating the supervector components $\mathcal{O}(\{n_{ij}\})$ with the help of a so-called \textit{generating function}:
\begin{equation}\label{eq:generating_function_coeffs}
    F(\{n_{ij}\})
    =
    \prod_{ \mathclap{ p, q = 1 } }^{M} \, \big( \, \lambda_{pq} \big)^{n_{pq}},
\end{equation}
parameterized with $M^2$ numbers $\lambda_{pq}$. Then, the components of operator \eqref{eq:K_body_operator} in the occupation-number basis are found by differentiating this function:
\begin{equation}\label{eq:K_body_vector}
    \mathcal{O}^{(K)}(\{n_{pq}\})
    =
    \sum_{\pmb{ij}} \mathcal{O}_{\pmb{ij}}^{(K)} \dfrac{\partial^{K} F(\{n_{pq}\})}{\partial \lambda_{i_1 j_1} \partial \lambda_{i_2 j_2} \ldots \partial\lambda_{i_K j_K}} \bigg|_{ \mathclap{ \hspace{1.05 cm} \substack{ \\\\ \lambda_{pq} = \delta_{pq} }}}.
\end{equation}

%%%%%%%%%%%%%%%%%%%%%%%%%%%%%%%%%%%%%%%%%%%%%%%%%
\subsection{Liouville master equation}\label{sec:III.C}
%%%%%%%%%%%%%%%%%%%%%%%%%%%%%%%%%%%%%%%%%%%%%%%%%
The density matrix evolves in time according to the master equation with a Liouville operator that is invariant under atomic permutations:
\begin{equation*}
    \dfrac{d\widehat{\rho}(t)}{dt}
    =
    \widehat{\mathcal{L}} \, \widehat{\rho}(t).
\end{equation*}
To apply our formalism, the Liouville operator should be bosonized following Section \ref{sec:II.C}. The initial condition has the form of eq.\ (\ref{eq:density_matrix_n}). To account for the temporal evolution, the expansion coefficients acquire an additional time argument: $\rho(\{n_{ij}\}, t)$.

We assume the Liouvillian to be described by Hamiltonian $\widehat{H}$ and an operator $\widehat{\mathcal{D}}$, which accounts for various incoherent processes, such as dephasing, pumping, damping, and others \cite{briegel1993quantum}:
\begin{equation}\label{eq:Liouvillian}
    \widehat{\mathcal{L}}
    =
    \dfrac{i}{\hbar} \left\{\widehat{H}^{\,\intercal} - \widehat{H} \right\} + \widehat{\mathcal{D}}.
\end{equation}
The generic trace-preserving dissipation operator has the Lindbladian form:
\begin{equation*}
    \begin{split}
        \widehat{\mathcal{D}} 
        = & 
        \sum_{\mathclap{i,j,p,q = 1}}^{M}\, \Gamma_{ijpq} \, \sum_{\mu = 1}^{N} \bigg\{ \widehat{\sigma}_{\mu, ij}\, \widehat{\sigma}^{\,\intercal}_{\mu, qp} - \dfrac{\delta_{ip}}{2} \bigg( \widehat{\sigma}_{\mu, qj} + \widehat{\sigma}_{\mu, qj}^{\,\intercal} \bigg) \bigg\} 
        \\
        = &
        \sum_{\mathclap{i,j,p,q = 1}}^{M} \, \Gamma_{ijpq} \, \bigg\{ \widehat{b}_{ip}^{\,\dagger} \, \widehat{b}_{jq}  - \dfrac{\delta_{ip}}{2} \sum_{t = 1}^{M} \bigg( \widehat{b}_{qt}^{\,\dagger} \, \widehat{b}_{jt} + \widehat{b}_{tj}^{\,\dagger} \, \widehat{b}_{tq} \bigg) \bigg\},
    \end{split}
\end{equation*}
with the coefficients $\Gamma_{ijpq}$ specifying the interaction of particles with a reservoir. These coefficients also ensure that the dissipator is a completely positive map \cite{Manzano2020}.

%%%%%%%%%%%%%%%%%%%%%%%%%%%%%%%%%%%%%%%%%%%%%%%%%
\subsection{Two-time correlation functions}\label{sec:III.D}
%%%%%%%%%%%%%%%%%%%%%%%%%%%%%%%%%%%%%%%%%%%%%%%%%

Some applications require the computation of two-time correlation functions in the form:
\begin{equation*}
    X(t, \tau)
    =
    \left\langle \widehat{\mathcal{A}}^{\,\dagger}(t + \tau) \, \widehat{\mathcal{B}}\,(t) \right\rangle = \text{Tr}\left\{ \widehat{\mathcal{A}}^{\,\dagger} \, \widehat{U}(\tau) \, \widehat{\mathcal{B}} \, \widehat\rho(t)\right\},
\end{equation*}
where $\widehat{\mathcal{A}}$ and $\widehat{\mathcal{B}}$ are atomic operators and $\widehat{U}(\tau)$ is an evolution operator
\begin{equation*}
    \frac{d\widehat{U}(t)}{dt}=\widehat{\mathcal{L}}\,\widehat{U}(t), \quad \widehat{U}(0)=1.
\end{equation*}
Within our framework, we interpret the operator $\widehat{\mathcal{A}}^{\,\dagger} \, \widehat{U}(\tau)$ as a supervector $ \llangle \mathcal{A} (\tau) |$ (see section \ref{sec:III.B} for more details). Consequently, the two-time correlation function has the following form:
\begin{equation}\label{eq:two_time}
    X(t, \tau)
    =
    \llangle \mathcal{A} (\tau) | \, \widehat{\mathcal{B}} \, |\rho(t) \rrangle.
\end{equation}
The evolution of this supervector is governed by the adjoint master equation \cite{Breuer2007}:
\begin{equation}\label{eq:adjoint_me}
     \dfrac{d |\mathcal{A}(\tau) \rrangle}{d\tau}
     =
     \widehat{\mathcal{L}}^{\,\dagger}  |\mathcal{A}(\tau) \rrangle. 
\end{equation}
Its solution gives the expression for the time-dependent object that is utilized for finding the specific correlation function. Note, in the absence of dissipation, equation\,\,(\ref{eq:adjoint_me}) transforms into a true Heisenberg equation.

%%%%%%%%%%%%%%%%%%%%%%%%%%%%%%%%%%%%%%%%%%%%%%%%%
\section{Applications of the formalism}\label{sec:IV}
%%%%%%%%%%%%%%%%%%%%%%%%%%%%%%%%%%%%%%%%%%%%%%%%%

To illustrate the capabilities of our method, we investigate the effect of dissipation and statistical mixing in the interaction between identical atoms and light. Numerical analysis was performed with the help of Julia and the DifferentialEquations.jl library \cite{Rackauckas2017}.

%%%%%%%%%%%%%%%%%%%%%%%%%%%%%%%%%%%%%%%%%%%%%%%%%
\subsection{Cooperative emission of two-level atoms}\label{sec:IV.A}
%%%%%%%%%%%%%%%%%%%%%%%%%%%%%%%%%%%%%%%%%%%%%%%%%

%%%%%%%%%%%%%%%%%%%%%%%%%%%%%%%%%%%%%%%%%%%%%%%%%%%%%%%%%%%%%%%%%

\begin{figure*}[t!]
    \centering
    \includegraphics[width = \linewidth]{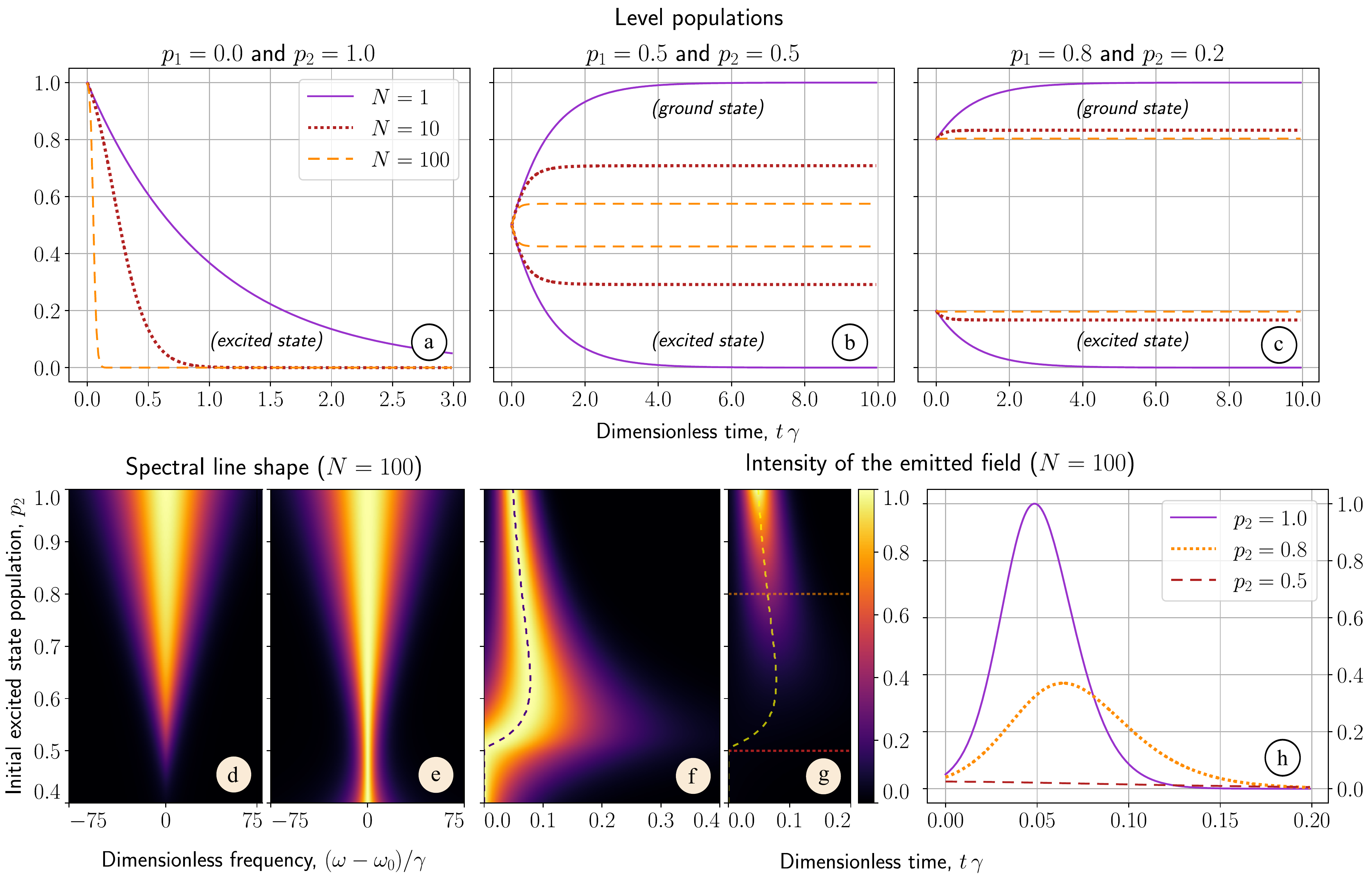}
    \caption{Evolution of the compact system of two-level atoms, initially prepared in a mixed state \eqref{eq:mixed_initial} with the initial probabilities (a) $p_1 = 0.0$ and $p_2 = 1.0$; (b) $p_1 = 0.5$ and $p_2 = 0.5$; (c) $p_1 = 0.8$ and $p_2 = 0.2$; calculated for the different number of particles $N$. (d) The spectral lineshape for varying initial conditions for $N = 100$, normalized to the overall maximum of distribution; (e) the same plot normalized for each row. (f) The intensity of the emitted field for different values of $p_2$, with each row normalized to the maximum of distribution; the dashed line represents the position of the peak intensity; (g) the same plot normalized to the overall maximum. (h) The intensity of the emitted field for specific values $p_2 = 1.0$, $0.8$, and $0.5$.}
    \label{fig:compact}
\end{figure*}

%%%%%%%%%%%%%%%%%%%%%%%%%%%%%%%%%%%%%%%%%%%%%%%%%%%%%%%%%%%%%%%%

As a first illustration, we study a compact system of atoms coupled to the bath of vacuum field modes. Despite its simplicity, this model predicts the fundamental phenomenon of coherent cooperative emission \cite{dicke1954coherence, gross1982superradiance}. The field degrees of freedom can be eliminated in the Born-Markov approximation \cite{ressayre1978markovian, Banfi1975, agarwal1974quantum, gross1982superradiance, carmichael1999statistical}. The resulting master equation in the rotating wave approximation is well-known:
\begin{equation*}
    \dfrac{d \widehat{\rho}(t)}{dt} = \widehat{\mathcal{L}}_\text{coll} \, \widehat{\rho}(t)
    =
    \gamma \, \widehat{J}_- \, \widehat{\rho}(t) \, \widehat{J}_+
    - \dfrac{\gamma}{2} \big[\widehat{\rho}(t), \,\widehat{J}_+ \, \widehat{J}_- \big]_+.
\end{equation*}
In our notation, $\widehat{J}_+ = \sum_\mu \widehat{\sigma}_{\mu, 21}$ and $\widehat{J}_-  = \sum_\mu \widehat{\sigma}_{\mu, 12}$, and $\gamma$ is the spontaneous decay rate. We neglect the dipole-dipole interactions that have a dephasing effect on the cooperative emission \cite{coffey1978effect, vogt2007electric, comparat2010dipole}. This approximation is valid for atoms in a cavity \cite{raimond1982statistics, hildred1984quantum} since non-resonant interactions are suppressed, and the dipole-dipole interaction is modified \cite{goldstein1997dipole}.

In the bosonized representation, the Liouvillian becomes:
\begin{equation}\label{eq:me_born_markov}
    \begin{split}
        \widehat{\mathcal{L}}_\text{coll}
        =
        \gamma \bigg\{ \widehat{b}_{11}^{\,\dagger} \, \widehat{b}_{22}  - \dfrac{1}{2} \sum_{t = 1}^{2} \left( \widehat{n}_{2t} + \widehat{n}_{t2} \right) \bigg\} \;
        \\
        - \dfrac{\gamma}{2} \, \sum_{\mathclap{p, q = 1}}^{2} \; \widehat{b}_{1q}^{\, \dagger} \left\{ \widehat{b}_{2p}^{\,\dagger} \, \widehat{b}_{1p} - \widehat{b}_{p1}^{\,\dagger} \, \widehat{b}_{p2} \right\} \, \widehat{b}_{2q} \;
        \\
        - \dfrac{\gamma}{2} \, \sum_{\mathclap{p,q = 1}}^{2} \; \widehat{b}^{\,\dagger}_{q1} \left\{ \widehat{b}^{\,\dagger}_{p2} \, \widehat{b}_{p1} - \widehat{b}^{\,\dagger}_{1p} \, \widehat{b}_{2p} \right\} \, \widehat{b}_{q2}.
    \end{split}
\end{equation}
The first line describes the independent decay of each atom, and the rest accounts for their collective interaction. The corresponding master equation in the occupation-number representation is written in Appendix \ref{App:me_A}.

Commonly, the atomic ensemble is assumed to be prepared in a fully inverted state. This assumption greatly simplifies the analysis, since the system is represented by a single value of the angular momentum $J = N/2$. Our method, however, allows treating all values of $J$, as well as permutation invariant statistical mixtures.

We assume our system to be prepared in a statistically mixed state, not decomposable in the basis of states with $J = N/2$:
\begin{equation}\label{eq:mixed_initial}
    \widehat{\rho}(0)
    =
    \prod_{\mu = 1}^{N} \left\{\sum_{i = 1}^{2}\, p_i \, \widehat{\sigma}_{\mu, ii}\right\},
\end{equation}
with $p_1 + p_2 = 1$. This state is translated to occupation numbers following Table \ref{table:1}. Within the Born-Markov approximation, the properties of the emitted field are expressed through atomic operators, which we discuss in Appendix \ref{Appendix:TLS_observables}.

The Liouvillian \eqref{eq:me_born_markov} commutes with the operator $\widehat{n}_{12} - \widehat{n}_{21}$. Since the system starts from the state \eqref{eq:mixed_initial}, only the density matrix elements with $n_{12} = n_{21} = \ell$ have nontrivial time evolution (see Appendix \ref{App:me_A}), leaving only
\begin{equation*}
\left[ \dfrac{(N+2)^2}{4} \right] \sim \dfrac{N^2}{4}
\end{equation*}
nontrivial elements, with $[\frac{a}{b}]$ denoting an integer division of $a$ and $b$.

When the system starts from a fully excited state $p_2 = 1.0$, the Liouvillian \eqref{eq:me_born_markov} describes superradiance. Fig.\! \ref{fig:compact}\! (a) shows the evolution of the excited state population for $N = 1$, $N = 10$ and $N = 100$. The latter exhibits much faster decay compared to the single-atom case, indicating the strong cooperative effect.

The picture changes drastically when the initial state is statistically mixed. Figs.\! \ref{fig:compact}\! (b) and (c) show level populations for different $N$ assuming $p_2 = 0.5$ and $p_2 = 0.2$, respectively. A single atom still obeys an exponential law and relaxes to the ground state. When $N > 1$, the excited state does not relax completely but reaches a nonzero steady value $p^{({ss})}_2$.

%%%%%%%%%%%%%%%%%%%%%%%%%%%%%%%%%%%%%%%%%%%%%%%%%%%%%%%%%%%%%

\begin{figure*}[t!]
    \centering
    \includegraphics[width=\linewidth]{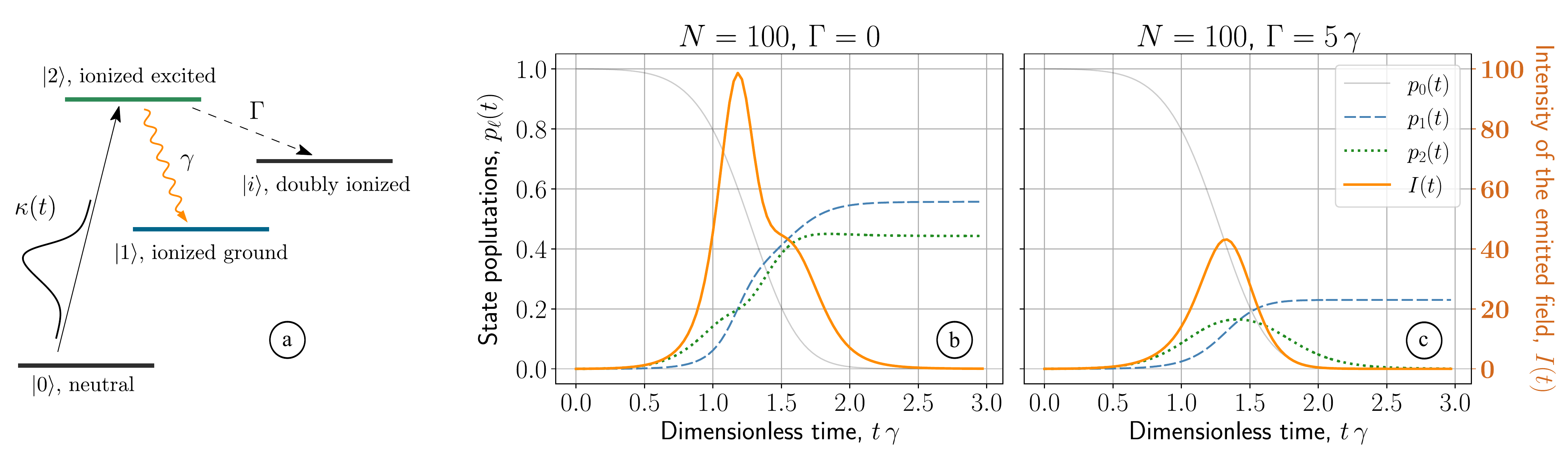}
    \caption{The evolution of the ensemble of $N=100$ atoms with the level structure depicted in (a). Neutral atoms are ionized by the photon flux with the Gaussian envelope $\kappa(t) = I_p \; e^{-\frac{(t-t_0)^2}{2\tau^2}} \big/ \sqrt{2\pi \tau^2}$, parameters used for calculations are $I_p = 10$, $t_0 = 2 \big/ \gamma$, $\tau = 0.5 \big/ \gamma$. Without the Auger effect, a well-pronounced superradiant burst is observed (b), which is significantly suppressed in the presence of fast Auger decay $\Gamma = 5 \gamma$ (c). The integral of the intensity yields the number of emitted photons.}
    \label{fig:lambda}
\end{figure*}

%%%%%%%%%%%%%%%%%%%%%%%%%%%%%%%%%%%%%%%%%%%%%%%%%%%%%%%%%%%%%

Fig.\! \ref{fig:compact}\! (d) shows the spectral line shape of the emission of $N=100$ atoms for different initial conditions. In Fig.\! \ref{fig:compact}\! (e) we normalize the spectrum for each row to demonstrate the narrowing due to the suppression of superradiance. Decreasing the initial excited state population also results in a temporal shift of the emission peak, which is signified by the vertical dashed line in Figs.\! \ref{fig:compact}\! (f) and \!(g). These plots depict the intensity of the emitted field for $N = 100$, normalized for each row (f) and normalized to the overall maximum (g). The intensities for selected values of $p_2$ are shown in Fig.\! \ref{fig:compact}\! (h).

Since the emission intensity is asymptotically zero, the appearance of steady populations indicates the presence of non-vanishing atomic correlations (see eq. \eqref{eq:intensity} of Appendix \ref{Appendix:TLS_observables}):
\begin{equation*}
  N \; p^{({ss})}_2
  =
  -  \sum_{\mathclap{n_{11} + 2 + n_{22} = N}} \; \; \rho^{({ss})}(\{n_{11}, 1, 1, n_{22} \}).
\end{equation*}
One can show that the field's intensity-intensity correlations also vanish with $t\rightarrow \infty$, i.e. there are nonzero higher-order atomic correlations that compensate the field observables. The analytical expression for the steady-state density matrix is the following:
\begin{multline*}
    \rho^{({ss})}(\{n_{11},\ell, \ell,n_{22}\}) 
    = 
    \dfrac{(n_{11} - n_{22} + 1)\, N!}{n_{22}!\, (n_{11} + \ell + 1)!\, \ell!} 
    \\
    \times \sum_{ \mathclap{ k=n_{22} + \ell} }^{[N/2]} \, (-1)^{k+n_{22}} \; (p_1 \, p_2)^k \; {{n_{11} + \ell -k}\choose{k - n_{22} - \ell}},
\end{multline*}
where $[N/2]$ denotes an integer division of $N$ and $2$. This expression defines the class of steady-states distinguished by the characteristic parameter $\left( p_1 \, p_2 \right)$. The analytical expression for the steady-state excited state population:
\begin{equation*}   
    p_2^{({ss})} = \sum_{k=1}^{[N/2]} C_{k-1} \; (p_1 \, p_2)^{k} \; \dfrac{N - 2k + 1}{N},
\end{equation*}
where $C_k = (2k)!\big/ k!\, (k+1)!$ are Catalan numbers. 

Qualitatively, the population trapping effect is interpreted as a dynamic balance between emission and absorption processes. However, in real physical systems, various dissipation channels and dipole-dipole interactions disrupt this balance, eventually leading to the complete relaxation of the whole ensemble.

Despite the relatively slow scaling of the number of equations $\sim N^2$, we encountered instabilities when solving $N\gtrsim 150$ with explicit methods. On the other hand, implicit methods are too computationally expensive for large systems of equations.

%%%%%%%%%%%%%%%%%%%%%%%%%%%%%%%%%%%%%%%%%%%%%%%%%%%%%%%%%%%%%%%
\subsection{Incoherently pumped three-level system}\label{sec:IV.B}
%%%%%%%%%%%%%%%%%%%%%%%%%%%%%%%%%%%%%%%%%%%%%%%%%%%%%%%%%%%%%%%

\begin{figure*}[t!]
    \centering
    \includegraphics[width = \linewidth]{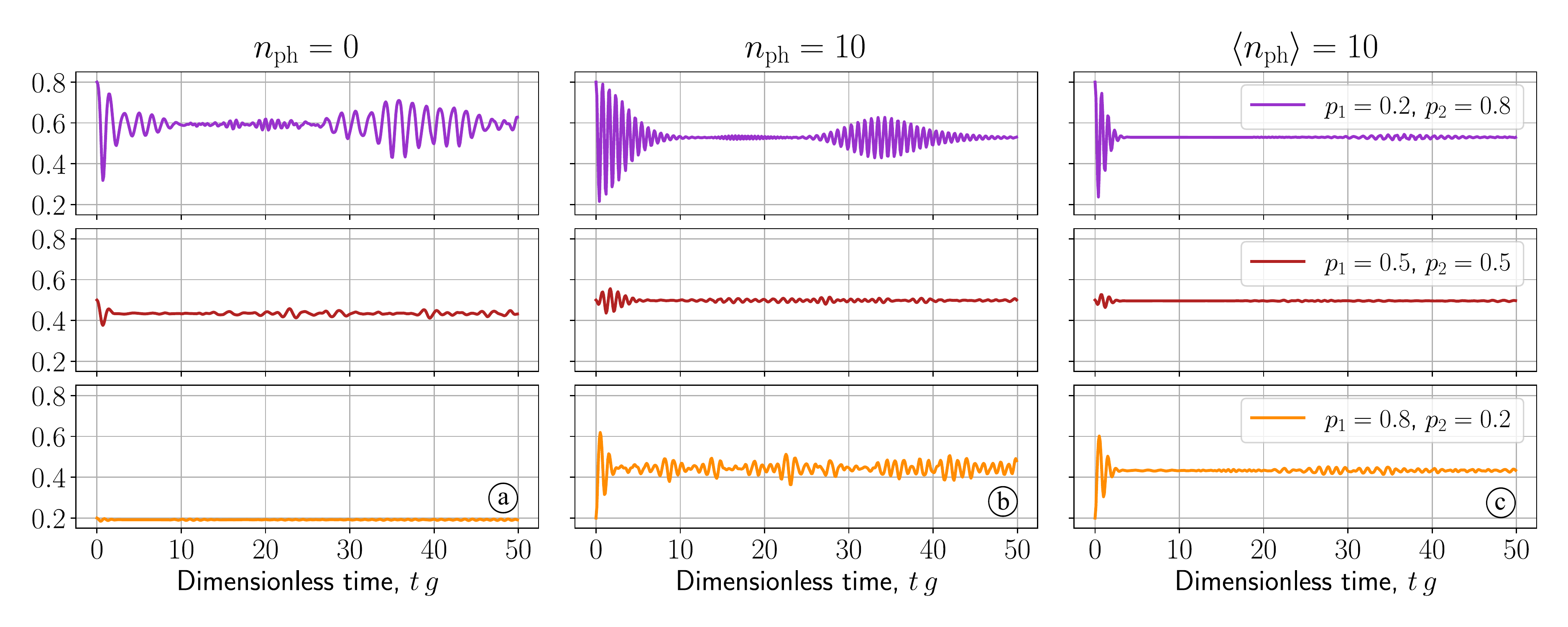}
    \caption{Probability to find an excited atom in the compact system of $N = 20$ atoms, initially prepared in a mixed state \eqref{eq:mixed_initial}, calculated for different initial conditions and initial field states: (a) vacuum state; (b) Fock state with $n_\mathrm{ph} = 10$ photons; (c) coherent state with $\langle n_\mathrm{ph} \rangle = 10$.}
    \label{fig:tcm}
\end{figure*}

A typically encountered experimental situation is an incoherently pumped $\Lambda$-system depicted in Fig.\! \ref{fig:lambda}\! (a), as in the simplified picture of inner-shell x-ray lasing experiments in neon \cite{Rohringer2012, Weninger2014}. Atoms in the neutral state $|0\rangle$ are ionized by an incoming pump pulse, treated as a flux of photons with the envelope $\kappa(t)$. The Lindbladian associated to this process reads:
\begin{equation*}
    \begin{split}
        \widehat{\mathcal{D}}_{0 \rightarrow 2}(t) = \kappa(t) \sum_{\mu=1}^{N} \left\{ \widehat{\sigma}_{\mu,20} \, \widehat{\sigma}^{\, \intercal}_{\mu,02} - \dfrac{1}{2} \left( \widehat{\sigma}_{\mu,00} + \widehat{\sigma}^{\, \intercal}_{\mu,00} \right) \right\} \,
        \\
        = \kappa(t) \left\{ \widehat{b}_{22}^{\,\dagger} \, \widehat{b}_{00} - \widehat{n}_{00} \right\}.
    \end{split}
\end{equation*}
The emerging excited inner-shell valence state $|2\rangle$ can decay by means of either the Auger effect to the doubly ionized state $|i\rangle$, according to the Lindblad operator:
\begin{multline*}
        \widehat{\mathcal{D}}_{2 \rightarrow i} = \Gamma
        \sum_{\mu=1}^{N} \left\{ \widehat{\sigma}_{\mu,i2} \, \widehat{\sigma}^{\, \intercal}_{\mu,2i} - \dfrac{1}{2} \left( \widehat{\sigma}_{\mu,22} + \widehat{\sigma}^{\, \intercal}_{\mu,22} \right) \right\} \,
        \\
         = \Gamma \left\{ \widehat{b}_{ii}^{\,\dagger} \, \widehat{b}_{22} - \dfrac{1}{2} \sum_{t = 1}^{2} \left( \widehat{n}_{2t} + \widehat{n}_{t2} \right) \right\},
\end{multline*}
or via cooperative radiative decay to the ground state $|1\rangle$, described by \eqref{eq:me_born_markov}. In total, the master equation consists of these three contributions:
\begin{equation*}
    \widehat{\mathcal{L}}(t)
    =
    \widehat{\mathcal{D}}_{0\rightarrow 2}(t) + \widehat{D}_{2\rightarrow i} + \widehat{\mathcal{L}}_\text{coll}.
\end{equation*}

The density matrix is expanded over the sets $\{n_{00}, n_{11}, n_{12}, n_{21}, n_{22}, n_{ii}\}$ with $n_{12} = n_{21} = \ell$. Two additional numbers $n_{00}$ and $n_{ii}$ count neutral atoms and doubly charged ions, respectively. Without the Auger process, i.e. $n_{ii} \equiv 0$ ($\Gamma = 0$), the size of basis scales as
\begin{equation*}
    \left[ \dfrac{(N+2) \, (N+4) \, (2N+3)}{24} \right] \sim \dfrac{N^3}{12},
\end{equation*}
while the inclusion of a doubly ionized state ($\Gamma \not = 0$) changes this to:
\begin{equation*}
    \left[ \dfrac{(N+2) \, (N+4) \, (N^2 + 6N+6)}{48} \right] \sim \dfrac{N^4}{48}.
\end{equation*}

In Fig.\! \ref{fig:lambda}\! (b), we see the dynamics without the Auger effect for $N = 100$ atoms. The pump flux is treated as a Gaussian with the width $\tau = 0.5 \big/ \gamma$ and the area $I_p = 10$. Similarly to the system from the previous section, this ensemble evolves into a steady state with a nonzero probability to find an excited atom. The inclusion of the Auger decay $\Gamma = 5 \, \gamma$ disrupts this balance and suppresses significantly the intensity of the emission, see Fig.\! \ref{fig:lambda}\! (c).

In a realistic system, the Auger decay rate $\Gamma$ exceeds the spontaneous decay rate $\gamma$ by a few orders of magnitude, and significantly more atoms are required to observe the cooperative emission. Unfortunately, even with the polynomial scaling $\sim N^4$ the computational cost is too high.

%%%%%%%%%%%%%%%%%%%%%%%%%%%%%%%%%%%%%%%%%%%%%%%%%%%%%%%%%%%%%%%
\subsection{Interaction with a single-mode field}\label{sec:IV.C}
%%%%%%%%%%%%%%%%%%%%%%%%%%%%%%%%%%%%%%%%%%%%%%%%%%%%%%%%%%%%%%%

In cavities, interaction with a finite set of field modes is predominant. Interaction of two-level atoms with the single-mode field is governed by \cite{tavis1968exact, garraway2011dicke, Larson2021, Hepp1973}:
\begin{equation}\label{eq:Hamiltonian}
     \widehat{V} =
     \hbar g \left( \widehat{J}_+ \, \widehat{a} + \widehat{a}^{ \dagger} \, \widehat{J}_- \right),
\end{equation}
where the rotating-wave approximation is applied. A suitable basis for composite density matrices consists of vectors
\begin{equation}\label{eq:tcm_basis}
    |\{n_{ij}\} \rrangle \otimes |n_L\rangle \langle n_R|,
\end{equation}
where $\ket{m}$ denotes the field Fock state with $m$ quanta. Expansion coefficients of the density matrix contain all possible information about the subsystems. Importantly, they describe the entanglement between them, implying that atomic and field degrees of freedom do not factorize.

After performing the second quantization, the Liouvillian \eqref{eq:Liouvillian} with the Hamiltonian \eqref{eq:Hamiltonian} becomes:
\begin{multline*}
        \widehat{\mathcal{L}}
        = i g \sum_{\mathclap{t = 1}}^{2} \! \bigg\{
        \widehat{a}^{\dagger \intercal} \, \widehat{b}_{t2}^{\,\dagger}\, \widehat{b}_{t1} - \widehat{a}^{\dagger}\, \widehat{b}_{1t}^{\,\dagger} \, \widehat{b}_{2t} 
        \\ + \widehat{b}_{t1}^{\,\dagger} \, \widehat{b}_{t2}\, \widehat{a}^{\intercal} - \widehat{b}_{2t}^{\,\dagger} \, \widehat{b}_{1t} \, \widehat{a} \bigg\}.
\end{multline*}
The explicit equation for the density matrix is given in Appendix \ref{App:me_B}. The evolution of each basis element \eqref{eq:tcm_basis} is characterized by two conservation laws at the level of occupation numbers, namely:
\begin{eqnarray*}
    (n_{22} - n_{11}) + (n_R + n_L) = \mathrm{const},\\
     (n_{12} - n_{21}) + (n_R - n_L)  = \mathrm{const},
\end{eqnarray*}
since corresponding operators commute with the Liouvillian. In a numerical sense, parallelization over the initial conditions allows taking advantage of these relations. It is important to note that the first relation may be violated if dissipation is included, e.g. the leakage of photons from the cavity. 

We assume atoms to be initially in the statistically mixed state \eqref{eq:mixed_initial}. The influence of the statistical mixing on the quantum dynamics is shown in Fig.\! \ref{fig:tcm} for different initial field states, where we plot atomic excited state population after tracing out the photonic subsystem. In the case of interaction with the vacuum $n_\text{ph} = 0$, instead of well-pronounced Rabi oscillations \cite{agarwal1985vacuum}, we see in Fig.\! \ref{fig:tcm}\! (a) the dynamics resembling the collapse phenomenon. Injecting $n_\text{ph} = 10$ photons in a Fock state results in more pronounced collapse-revivals for $p_1 = 0.2$, $p_2 = 0.8$, as depicted in Fig.\! \ref{fig:tcm}\! (b). The revival effect is weaker and collapse occurs earlier when the field is initially in the coherent state with $\langle n_\text{ph} \rangle = 10$, as plotted in Fig.\! \ref{fig:tcm}\! (c).

%%%%%%%%%%%%%%%%%%%%%%%%%%%%%%%%%%%%%%%%%%%%%%%%%
\section{Outlook on phase-space description}\label{sec:V}
%%%%%%%%%%%%%%%%%%%%%%%%%%%%%%%%%%%%%%%%%%%%%%%%%

In our numerical illustrations, we considered medium-sized quantum systems. When a large number of particles is involved ($N \gtrsim 10^3)$, even the polynomial scaling limits numerical implementation. Phase-space techniques are capable of resolving this computational bottleneck. These methods rely on a conversion of the quantum master equation into a Fokker-Planck equation for appropriate phase-space distribution functions \cite{drummond2014quantum, carmichael1999statistical, drummond1980generalised, scully1999quantum}. It is well-formulated for bosonic systems in the coherent state basis, where the density matrix is represented as a combination of the following projectors \cite{drummond2014quantum}:
\begin{equation}\label{eq:lambda_coherent}
    \widehat{\Lambda}(\alpha, \alpha^\dagger) = \dfrac{|\alpha\rangle \langle \alpha^{\dagger *}|}{\langle \alpha^{\dagger *} | \alpha \rangle },
\end{equation}
and can be extended to atomic systems \cite{drummond1991quantum, olsen2005phase, Huber2021, Mandt2015, narducci1978transient, puri1979exact}, e.g. by introducing spin-coherent states \cite{barry2008qubit, arecchi1973coherent, radcliffe1971some, arecchi1972atomic}. The projectors \eqref{eq:lambda_coherent} transform bosonic operators into $c$-numbers, leading to the property
\begin{equation}\label{eq:lambda_property}
    \mathrm{Tr} \, \bigg\{ \widehat{a}^{ \dagger} \, \widehat{a} \; \widehat{\Lambda}(\alpha, \alpha^\dagger) \bigg\} = \alpha^\dagger  \alpha,
\end{equation}
namely, coherent ket-vector turns $\widehat{a}$ into $\alpha$, whereas coherent bra-vector turns $\widehat{a}^\dagger$ into $\alpha^\dagger$.

To merge this technique with our second-quantization formalism, we introduce a similar ansatz for the atomic density matrix:
\begin{equation*}
    \widehat{\rho}(t)
    = 
    \int \prod_{\mathclap{i,j=1}}^M d^2 \beta_{ij} \, P(\{ \beta_{ij} \}, t) \; \widehat{\Lambda}( \beta_{ij}),
\end{equation*}
where $\widehat{\Lambda}( \beta_{ij})$ is an operator analogous to \eqref{eq:lambda_coherent}, but related to coherent states for bosonic superoperators $\widehat{b}_{ij}$. The integration runs over complex numbers $\beta_{ij}$ which have the same role as $\alpha, \alpha^\dagger$ in \eqref{eq:lambda_coherent}.

Being the element of the Liouville-Hilbert space, the operator $\widehat{\Lambda}( \beta_{ij})$ is isomorphic to some supervector $|\Lambda( \beta_{ij})\rrangle$. For example, for $\sigma$-operators the following holds:
\begin{equation}
\label{eq:lambda_operator:sigma_into_b}
    \sum_{\mu = 1}^{N} \mathrm{Tr} \left\{ \widehat{\sigma}_{\mu,pq} \, \widehat{\Lambda}(\beta_{ij} ) \right\}
    \equiv
    \mathrm{Tr} \left\{ \widehat{b}^{\, \dagger}_{pp} \, \widehat{b}_{qp} \; |{\Lambda}(\beta_{ij} ) \rrangle \right\}.
\end{equation}
To obtain a similar property as in \eqref{eq:lambda_property}, one could be tempted to construct the supervector $|\Lambda( \beta_{ij})\rrangle$ out of coherent states for $b$-operators. This would turn $\widehat{b}_{qp}$ into $\beta_{qp}$, however, $\widehat{b}^{\, \dagger}_{pp}$ would not lead to a simple expression, since we cannot insert a bra-vector in supervector $|\Lambda( \beta_{ij})\rrangle$ similarly to $\langle \alpha^{\dagger *}|$ in eq. \eqref{eq:lambda_coherent}. Fortunately, a slight modification of the coherent state ket-vector gives the desired properties:
\begin{equation*}
    | \Lambda(\beta_{ij})\rrangle = \exp\left(- \sum_{i=1}^{M} \beta_{ii} + \dfrac{1}{\sqrt{\widehat{N}}} \sum_{\mathclap{i,j=1}}^{M} \beta_{ij} \, \widehat{b}_{ij}^{\, \dagger} \right) |\mathrm{vac}\rrangle.
\end{equation*}
The first term in the exponent ensures that this state has a unit trace. The only difference with the usual coherent states is the additional operator $1/\sqrt{\widehat{N}}$. Similarly to the traditional $\Lambda$-projectors, it turns the product $\widehat{b}_{pq}\sqrt{\widehat{N}}$ into $\beta_{pq}$
\begin{equation*}
    \widehat{b}_{pq}\sqrt{\widehat{N}} \, | \Lambda(\beta_{ij})\rrangle
    =
    \beta_{pq} \, |\Lambda(\beta_{ij})\rrangle.
\end{equation*}
Besides, it conveniently simplifies eq. \eqref{eq:lambda_operator:sigma_into_b}:
\begin{equation*}
    \sum_{\mu = 1}^{N} \mathrm{Tr} \left\{ \widehat{\sigma}_{\mu,pq} \, \widehat{\Lambda}(\beta_{ij} ) \right\} =\beta_{qp}.
\end{equation*}

Consequently, $\beta$-numbers directly represent $\sigma$-operators at the level of the Fokker-Planck equation. At the same time, the quantum master equation translates into the equation for the $P$-function using the replacements:
\begin{align*}
&\widehat{b}_{ij} \, | \rho \rrangle \rightarrow \beta_{ij} \, P,
& \widehat{b}^{\,\dagger}_{ij} \, | \rho \rrangle \rightarrow \left(\delta_{ij} - \dfrac{\partial}{\partial \beta_{ij}} \right) \, P.
\end{align*}
If the Liouvillian contains only the terms with at most two creation and two annihilation operators, the resulting equation has the form of the Fokker-Planck equation.

%%%%%%%%%%%%%%%%%%%%%%%%%%%%%%%%%%%%%%
\section{Conclusion}
%%%%%%%%%%%%%%%%%%%%%%%%%%%%%%%%%%%%%%

The main result of our work is the formulation of the second quantization of density matrices, which incorporates incoherent processes into this framework. Due to this, the physical intuition of second quantization in Hilbert space applies to Liouville space and density matrices. The number of degrees of freedom reduces from exponential to polynomial scaling with the number of particles. Our formalism describes particles with arbitrary discrete level structures and assumes conservation of the number of particles --- a non-relativistic limit. We studied the effects of statistical mixing and dissipation in the context of light-matter interactions for systems of up to $N = 100$ emitters. To overcome the numerical limitations with respect to the particle number, we formulated the link to an appropriate phase-space technique.

\begin{acknowledgments}

We would like to thank Dr. Aliaksandr Leonau for the constructive and critical discussion of the manuscript. We acknowledge the financial support of Grant-No.\! HIDSS-0002 DASHH (Data Science in Hamburg-Helmholtz Graduate School for the Structure of Matter).

\end{acknowledgments}

\bibliography{apssamp}% Produces the bibliography via BibTeX.

%%%%%%%%%%%%%%%%%%%%%%%%%%%%%%%%%%%%%%%%%%%%%%%%%%%%%%%%
%%%%%%%%%%%%%%%%%%%%%%%%%%%%%%%%%%%%%%%%%%%%%%%%%%%%%%%%

\appendix

%%%%%%%%%%%%%%%%%%%%%%%%%%%%%%%%%%%%%%%%%%%%
\section{Second quantization of $M$-level atoms}\label{Appendix:A}
%%%%%%%%%%%%%%%%%%%%%%%%%%%%%%%%%%%%%%%%%%%%

We exploit the permutation symmetry by disregarding the state of each individual particle and only count the number of particles $n_q$ in each state $|q\rangle$ with $q = 1, \ldots, M$. The description of the ensemble builds on the collective operators
\begin{equation}\label{eq:collective_J}
    \widehat{J}_{pq} = \sum_{\mu = 1}^{N} \widehat{\sigma}_{\mu, pq},
\end{equation}
which preserve the permutation symmetry. $\widehat{J}_{pq}$ annihilates a particle in a state $|q\rangle$ and creates one in a state $|p\rangle$. This operator can be represented by collective bosonic operators \cite{landau1977}:
\begin{equation}\label{eq:bosonization_Hilbert}
    \sum_{\mu = 1}^{N} \widehat{\sigma}_{\mu, pq} = \sum_{\mathclap{t, s = 1}}^{M} 
    \underbrace{\bra{s} \widehat{\sigma}_{pq} \ket{t}}_{\delta_{sp} \, \delta_{qt}} \, \widehat{b}_s^{\,\dagger}\, \widehat{b}_t = \widehat{b}_p^{\,\dagger}\, \widehat{b}_q,
\end{equation}
with $\big[ \, \widehat{b}_p, \widehat{b}_q^{\,\dagger}\big] = \delta_{pq}$. Symmetric states are created by acting on the vacuum state $|\text{vac}\rangle$:
\begin{equation}\label{eq:HS_occupation_numbers}
    |\{n_i\} \rangle = \prod_{q = 1}^{M} \dfrac{\big(\, \widehat{b}_q^{\,\dagger} \big)^{n_{q}} }{\sqrt{n_q!}} \, |\text{vac}\rangle,
\end{equation}
where $\{n_i\} = \{n_1, n_2, \ldots, n_M\}$ is the set of occupation numbers for each atomic state. 

From an algebraic point of view, the bosonization \eqref{eq:bosonization_Hilbert} is a particular case of the Jordan map \cite{biedenharn1984angular, klein1991boson}. Mapping $\widehat{J}_{ij}$ to bosonic operators according to eq.\! \eqref{eq:bosonization_Hilbert} preserves the commutation relations:
\begin{equation}\label{eq:J_algebra}
    \big[\widehat{J}_{ij}, \, \widehat{J}_{pq} \big] = \delta_{pj} \, \widehat{J}_{iq} - \delta_{iq} \, \widehat{J}_{pj}.
\end{equation}

%%%%%%%%%%%%%%%%%%%%%%%%%%%%%%%%%%%%%%%%%%%%%
\section{Derivation details}
%%%%%%%%%%%%%%%%%%%%%%%%%%%%%%%%%%%%%%%%%%%%%

%%%%%%%%%%%%%%%%%%%%%%%%%%%%%%%%%%%%%%%%%%%%%
\subsection{Left-acting operators}\label{App:B.1}
%%%%%%%%%%%%%%%%%%%%%%%%%%%%%%%%%%%%%%%%%%%%%

To find how $\sigma$-operators transform the basis vectors \eqref{eq:superket}, let us write their action on the density matrix \eqref{eq:density_matrix_sigma}:
\begin{equation*}
    \widehat{\sigma}_{\mu, pq} \; \widehat{\rho}
    = \sum_{\pmb{ij}} \overline{C
    _{\pmb{ij}}} \; \widehat{\sigma}_{1, i_1 j_1} \ldots \boxed{\delta_{qi_\mu} \, \widehat{\sigma}_{\mu,p j_\mu}} \ldots \widehat{\sigma}_{N, i_N j_N}.
\end{equation*}
In the vector representation \eqref{eq:density_matrix}, this reads
\begin{equation*}
    \widehat{\sigma}_{\mu, pq} \, | \rho \rrangle 
     = \sum_{\pmb{ij}} \overline{C
    _{\pmb{ij}}} \; |i_1 j_1 \rrangle_1 \ldots \boxed{\widehat{\sigma}_{\mu, pq} \, |i_\mu j_\mu \rrangle_{\mu}} \ldots|i_N j_N \rrangle_N.
\end{equation*}
And we immediately find:
\begin{equation*}
    \widehat{\sigma}_{\mu, pq} \, |ij\rrangle_\mu = \delta_{qi} \, |pj\rrangle_\mu.
\end{equation*}
Collective operators are quantized analogously to eq.\ \eqref{eq:bosonization_Hilbert}:
\begin{equation*}
    \sum_{\mu = 1}^{N}\, \widehat{\sigma}_{\mu,pq} 
    = 
    \mathlarger{\sum}\limits_{\mathclap{s,t,i,j = 1}}^{M}\, \underbrace{\llangle st| \, \widehat{\sigma}_{pq} \, |ij \rrangle }_{\delta_{sp} \, \delta_{qi} \, \delta_{jt}} \, \widehat{b}_{st}^{\,\dagger} \, \widehat{b}_{ij}
    = 
    \sum_{t = 1}^{M}\, \widehat{b}_{pt}^{\,\dagger} \, \widehat{b}_{qt}.
\end{equation*}

%%%%%%%%%%%%%%%%%%%%%%%%%%%%%%%%%%%%%%%%%%%%%%%%%%%%%%%
\subsection{Right-acting operators}\label{App:B.2}
%%%%%%%%%%%%%%%%%%%%%%%%%%%%%%%%%%%%%%%%%%%%%%%%%%%%%%%

As for the right-acting operators, they transform the density matrix \eqref{eq:density_matrix_sigma} as:
\begin{equation*}
    \widehat{\rho}\; \widehat{\sigma}_{\mu, k\ell}
     = \sum_{\pmb{ij}} \overline{C
    _{\pmb{ij}}} \; \widehat{\sigma}_{1, i_1 j_1} \ldots \boxed{\delta_{j_\mu k} \, \widehat{\sigma}_{\mu, i_\mu \ell} } \ldots \widehat{\sigma}_{N, i_N j_N}.
\end{equation*}
As pointed out in the main text, we denote the right-action with a superscript $^\intercal$:
\begin{equation*}
    \widehat{\sigma}^{\,\intercal}_{\mu, k\ell} \, |\rho \rrangle 
     = \sum_{\pmb{ij}} \overline{C_{\pmb{ij}}} \; |i_1 j_1\rrangle_1 \ldots \boxed{\widehat{\sigma}^{\,\intercal}_{\mu, k\ell} \, |i_\mu j_\mu\rrangle_\mu} \ldots|i_N j_N\rrangle_N,
\end{equation*}
which immediately leads to the relation:
\begin{equation*}
     \widehat{\sigma}_{\mu, k\ell}^{\, \intercal} \, |ij\rrangle_\mu = \delta_{jk} \, |i\ell\rrangle_\mu.
\end{equation*}
For the collective right-acting operator we thus get:
\begin{equation*}
    \sum_{\mu = 1}^{N} \widehat{\sigma}^{\,\intercal}_{\mu,k\ell} 
    =
    \mathlarger{\sum}_{\mathclap{s, t, i, j = 1}}^{M}\, 
    \underbrace{\llangle st|\, \widehat{\sigma}_{k\ell}^{\,\intercal} \, |ij\rrangle}_{\delta_{si} \, \delta_{jk} \, \delta_{\ell t}}
    \, \widehat{b}_{st}^{\,\dagger}\, \widehat{b}_{ij} 
    =
    \sum_{s = 1}^{M}\, \widehat{b}_{s\ell}^{\,\dagger} \, \widehat{b}_{sk}.
\end{equation*}

\section{Generating functional}\label{appendix:generating_functional}
%%%%%%%%%%%%%%%%%%%%%%%%%%%%%%%%%%%%%%%%%%%%%%%%

We call the following $N$-particle symmetric operator the \textit{generating functional}
\begin{equation}\label{eq:generating_functional}
    \widehat{F}(\lambda) = \prod_{\mu = 1}^{N} \left\{ \, \sum_{\mathclap{i, j = 1}}^{M} \lambda_{ij}\, \widehat{\sigma}_{\mu, ij} \right\},
\end{equation}
which is parameterized with $M^2$ numbers $\{\lambda_{ij}\}$. When differentiating this functional with respect to $\lambda_{pq}$, the symmetrized combination of $\sigma$-operators is picked out of the product:
\begin{equation*}
    \dfrac{\partial}{\partial \lambda_{pq}} \widehat{F}(\lambda) = \sum_{\mathclap{\nu=1}}^{N} \left( \widehat{\sigma}_{\nu,pq} \prod_{\substack{\mu = 1 \\ \mu \not = \nu}}^{N} \left\{ \, \sum_{\mathclap{i, j = 1}}^{M} \lambda_{ij}\, \widehat{\sigma}_{\mu, ij} \right\} \right).
\end{equation*}
Setting $\lambda_{ij} = \delta_{ij}$, the remaining sums under the $\prod$-sign become identities, and only the symmetrical operator $\widehat{J}_{pq}$ (see eq. \eqref{eq:collective_J}) is left. $K$-particle symmetrized operators require $K$-th order partial derivative. 

If we interpret \eqref{eq:generating_functional} as the non-normalized pure uncorrelated state from Table \ref{table:1} (by putting $c_i \, c_j^* \rightarrow \lambda_{ij}$), the occupation-number representation of the generating functional follows from the corresponding row  of the table:
\begin{equation}\label{eq:generating_functional_occupation}
    |F(\lambda) \rrangle = \sum_{\{m_{ij}\}} \sqrt{\dfrac{N!}{\prod_{pq} m_{pq}!} } \; F(\{m_{ij}\}) \, |\{m_{ij}\}\rrangle,
\end{equation}
where the coefficients $F(\{m_{ij}\})$ are given by
\begin{equation}
    F(\{m_{ij}\}) = \prod_{\mathclap{p, q = 1}}^{M} \, \big( \, \lambda_{pq} \big)^{m_{pq}}.
\end{equation}
By appropriately applying derivatives, it is possible to translate symmetrical operators into supervectors.

The \textit{generating function} for expectation values, also parameterized with the set $\{\lambda_{ij}\}$, is obtained by projecting density matrix \eqref{eq:density_matrix_n} onto the supervector \eqref{eq:generating_functional_occupation}, i.e. by averaging the functional \eqref{eq:generating_functional}:
\begin{equation}\label{eq:generating_function}
    \langle F (\lambda) \rangle = \sum_{\mathclap{\{n_{ij}\}}} \; \rho(\{n_{ji}\}) \, \prod_{\mathclap{p, q = 1}}^{M} \, \big( \, \lambda_{pq} \big)^{n_{pq}}.
\end{equation}
Differentiating the generating function gives averages of corresponding operators. 

%%%%%%%%%%%%%%%%%%%%%%%%%%%%%%%%%%%%%%%%%%%%%%%%%%%%%%%%%%%%%%%%%
\section{Observables of two-level systems}\label{Appendix:TLS_observables}
%%%%%%%%%%%%%%%%%%%%%%%%%%%%%%%%%%%%%%%%%%%%%%%%%%%%%%%%%%%%%%%%%

Although the field is seemingly absent in the master equation, its properties are encoded in appropriate atomic correlation functions. We are interested in the temporal and spectral intensities of the field. But first let us consider the one-particle operator $\widehat{J}_{qq}$ (see eq. \eqref{eq:collective_J}), which counts atoms in a state $q$ and is quantized according to eq.\ \eqref{eq:sigma_L_bosonization}.
Its expectation value divided by $N$ is the probability to find an atom in the state $q=1,2$:
\begin{equation*}
    p_q(t)
    =
    \dfrac{\mathrm{Tr}\big\{ \widehat{J}_{qq}\, \widehat{\rho}(t) \big\}}{N} =  \mathlarger{\sum}_{\mathclap{ \substack{\\[0.5mm] n_{11} + n_{22} = N }}}\;\, \dfrac{n_{qq}}{N} \; \rho(\{n_{11}, 0, 0, n_{22}\}, t).
\end{equation*}

The correlation properties of the total atomic dipole moment define the intensity of the emission into all spatial angles \cite{Benedict2018}:
\begin{equation*}
    \begin{split}
        \widehat{I}
        =
        \widehat{J}_{21} \, \widehat{J}_{12} = 
        \widehat{J}_{22} + \sum_{\mathclap{p, q = 1}}^{2} \, \widehat{b}_{2p}^{\, \dagger} \, \widehat{b}_{1q}^{\, \dagger} \, \widehat{b}_{1p} \, \widehat{b}_{2q}.
    \end{split}
\end{equation*}
We split this operator into two contributions: the number of excited atoms and the two-particle operator, which is quantized according to eq.\ \eqref{eq:two_body_quantization}. In the second term only the contributions $n_{12} = n_{21} = 1$ survive after averaging:
\begin{equation}\label{eq:intensity}
    \mathrm{Tr}\big\{ \widehat{I}\, \widehat{\rho}(t) \big\}
    =
    N \, p_2(t)  + \mathlarger{\sum}_{\mathclap{n_{11} + 2 + n_{22} = N}} \; \; \rho(\{n_{11}, 1, 1, n_{22}\}, t).
\end{equation}
This expectation value gives the temporal shape of the emission. As for the spectral line shape \cite{scully1999quantum}
\begin{equation*}
    S(\omega)
    \sim
    \mathrm{Re} \int_{0}^{\infty} \!\!  dt \int_{0}^{\infty} \!\! d\tau \, X(t, \tau)\, e^{- i \omega \tau},
    \end{equation*}
where $X(t, \tau)$ is the two-time correlation function:
\begin{equation*}
    X(t, \tau)
    =
    \langle \widehat{J}_{21}(t + \tau) \, \widehat{J}_{12}(t) \rangle = \llangle J_{12}(\tau) | \, \widehat{J}_{12} \, |\rho(t)\rrangle,
\end{equation*}
where the last equality is due to eq.\ \eqref{eq:two_time}. This average is calculated as described in Sec. \ref{sec:III.B}. The operator $\widehat{J}_{12}$ is propagated in time as a supervector according to the adjoint master equation \eqref{eq:adjoint_me} with the initial condition:
\begin{equation*}
        |J_{12}(0)\rrangle
        =
        \mathlarger{\sum}_{\mathclap{\substack{\\[0.5mm] n_{11} + 1 + n_{22} = N}}} \; \; \sqrt{\dfrac{N!}{n_{11}! \, n_{22}!} } \, |\{n_{11}, 1, 0, n_{22}\} \rrangle.
\end{equation*}
This representation is a direct consequence of eq.\ \eqref{eq:K_body_vector}.

\section{Master equations}

\subsection{Cooperative emission}\label{App:me_A}

The evolution of the density matrix \eqref{eq:density_matrix_n} according to the Liouvillian \eqref{eq:me_born_markov} is governed by the system of linear differential equations:
\begin{widetext}
\begin{equation*}
\begin{split}
    \dot{\rho}\big(\{n_{11}, n_{12}, n_{21}, n_{22}\},\tau\big) = -
    \left\{ n_{22} + \dfrac{n_{21} + n_{12}}{2} \; (n_{11}+n_{22} + 1) \right\} \; \rho\big(\{n_{11}, n_{12}, n_{21}, n_{22}\}, \tau \big) \;\\
    + (n_{22} + 1) (n_{12} + n_{21} + 1) \; \rho\big(\{n_{11}-1,n_{12}, n_{21}, n_{22} + 1\}, \tau \big) \;
    \\
    + (n_{12} + 1) (n_{21} + 1) \bigg\{ \rho\big(\{n_{11} - 2, n_{12} + 1, n_{21} + 1, n_{22}\}, \tau \big) -  \rho\big(\{ n_{11} - 1, n_{12} + 1, n_{21} + 1, n_{22} - 1\}, \tau \big) \bigg\}\;
    \\ 
     +  (n_{22}+1) \bigg\{ (n_{22} + 2) \, \rho\big(\{n_{11}, n_{12} - 1, n_{21} - 1, n_{22} + 2\}, \tau \big) -(n_{11} + 1) \, \rho\big(\{ n_{11} + 1, n_{12} - 1, n_{21} - 1, n_{22} + 1\}, \tau \big)  \bigg\},
\end{split}
\end{equation*}
\end{widetext}
where $\tau = t\, \gamma$. Numbers $n_{12}$ and $n_{21}$ change simultaneously by the same value. It means that the evolution of a single vector $|\{n_{11}, n_{12}, n_{21}, n_{22} \} \rrangle$, or the corresponding coefficient ${\rho}\big(\{n_{11}, n_{12}, n_{21}, n_{22}\},\tau\big)$, is characterised by the conservation law $n_{12} - n_{21} = \mathrm{const}$. In the state \eqref{eq:mixed_initial}, this constant is zero for each coefficient, implying nontrivial evolution only for $n_{12} = n_{21} = \ell$.

\subsection{Tavis-Cummings model}\label{App:me_B}

The decomposition coefficients of density matrix in the basis \eqref{eq:tcm_basis} denoted as $\rho\big( \{n_{11}, n_{12}, n_{21}, n_{22}, n_L, n_R\}, t\big)$ satisfy the equations, with $\tau = t \, g$:
\begin{widetext}
\begin{equation*}
\begin{split}
\dot{\rho}\big(\{n_{11}, n_{12}, n_{21}, n_{22}, n_L, n_R\},\tau\big) \; \;
\\ 
= i \sqrt{n_R + 1} \bigg\{ (n_{21} + 1) \, \rho\big(\{n_{11}, n_{12}, n_{21} + 1, n_{22} - 1, n_L, n_R + 1\}, \tau\big) + (n_{11} + 1) \, \rho(\{n_{11} + 1, n_{12} - 1, n_{21}, n_{22}, n_L, n_R + 1\}, \tau)\bigg\} \;
\\
- i\sqrt{n_L + 1} \bigg\{  (n_{12} + 1)  \, \rho\big(\{n_{11}, n_{12} + 1, n_{21}, n_{22} - 1, n_L + 1, n_R\}, \tau\big)
+ (n_{11} + 1)  \, \rho\big(\{n_{11} + 1, n_{12}, n_{21} - 1, n_{22}, n_L + 1, n_R\}, \tau\big)  \bigg\} \;
\\
+ i \sqrt{n_R} \bigg\{ (n_{12} + 1) \, \, \rho\big(\{n_{11} - 1, n_{12} + 1, n_{21}, n_{22}, n_L, n_R - 1\}, \tau\big) 
+ (n_{22} + 1) \, \rho\big(\{n_{11}, n_{12}, n_{21} - 1, n_{22} + 1, n_L, n_R - 1\}, \tau\big) \bigg\} \;
\\
- i \sqrt{n_L} \bigg\{ (n_{21} + 1) \, \, \rho\big(\{n_{11} - 1, n_{12}, n_{21} + 1, n_{22}, n_L - 1, n_R\}, \tau\big) 
+ (n_{22} + 1) \, \rho\big(\{n_{11}, n_{12} - 1, n_{21}, n_{22} + 1, n_L - 1, n_R\}, \tau\big) \bigg\}.
\end{split}
\end{equation*}
\end{widetext}

\end{document}